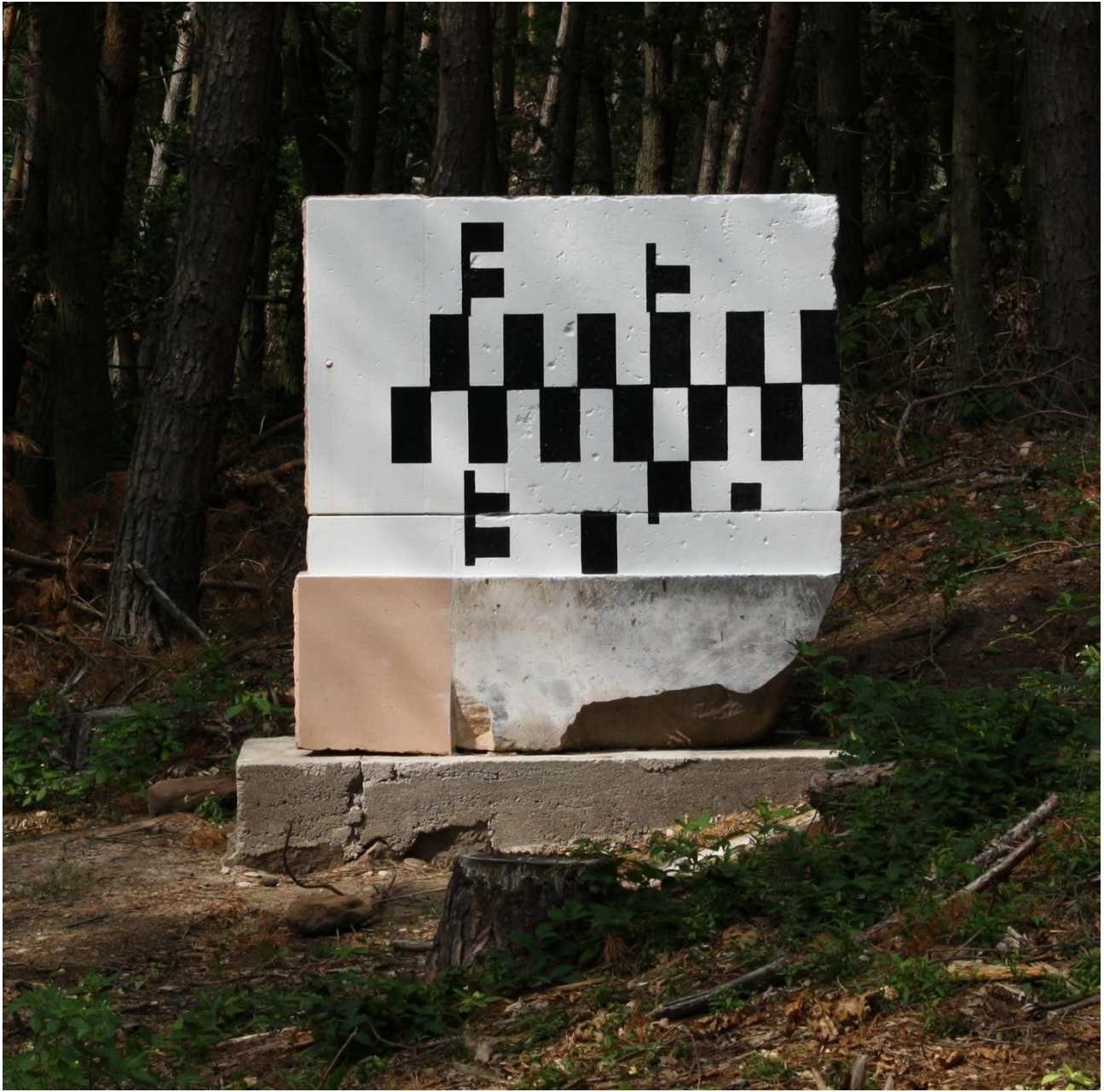

# Gerling-Sternwarte

Nach der Kopernikanischen Wende, der die Philippina immerhin eine Bewerbung von Giordano Bruno (1548–1600) auf einen ihrer Lehrstühle verdankt, und den frühen Entdeckungen am Sternenhimmel Anfang des 17. Jahrhunderts folgte eine Zeit der Ernüchterung für die Astronomie, da die damaligen Fernrohre offensichtlich außer einer genaueren Beobachtung der bekannten Planeten unseres Sonnensystems und der Sonne selber keine weiteren Entdeckungen mehr zuließen. Erst Ende des 18. Jahrhunderts gelang Wilhelm Herschel (1738–1822) mit seinen neuen Spiegelfernrohren die Entdeckung des Planeten Uranus. Auf seinem Grabstein heißt es: „Er durchbrach die Grenzen des Himmels". Die Astronomie erlebte einen Aufschwung und der Wunsch nach einer Sternwarte erreichte auch Marburg. So machte Johann Gottlieb Waldin (1728–1795), seit 1766 Ordinarius für Philosophie und Mathematik an der Philipps-Universität, den Vorschlag, zwei alte Pulvertürme auf dem Schlossberg zu einer Sternwarte umzurüsten. Anfang des 19. Jahrhunderts gab es während des kurzen Königreichs Westfalen Pläne für ein neues Bibliotheksgebäude mit einer Sternwarte. Beide Pläne konnten nicht realisiert werden.

Mit der Berufung von Christian Ludwig Gerling (1788–1864) als Professor für Mathematik, Physik und Astronomie erfolgte 1817 die Gründung des Mathematisch-Physikalischen Instituts der Philipps-Universität. Gerling hatte bei Carl Friedrich Gauß (1777–1855) promoviert und stand ganz in der astronomisch-mathematischen Tradition seines großen Lehrers und langjährigen Freundes. Das Institut war im Deutschen Haus untergebracht, welches sich bald als unzulänglich für eine physikalische und astronomische Ausbildung erwies. Nach langwierigem Ringen erhielt Gerling 1838 endlich die Zusage für neue, vor allem größere Räumlichkeiten im Hauptgebäude des ehemaligen Dörnberger Hofes am Renthof. Dort ließ er nicht nur die physikalische Sammlung – die für die Ausbildung in experimenteller Physik notwendigen Geräte – großzügig in einzelnen zugeordneten Zimmern fest und dauerhaft aufstellen, sondern baute den oberen Teil des alten Turms zu einer Sternwarte um. Im Jahre 1841 waren die Baumaßnahmen abgeschlossen, und er konnte die erste und einzige Sternwarte der Philipps-Universität in Betrieb nehmen.

**gegründet 1841**
**Sammlung:** Historische Sternwarte im Turm des alten Institutsgebäudes, Messsäulen in Galerie, Ausstellungsraum (ehemaliger Geräteraum), Meridianstein in Wehrda, Oststein am Ortenberg

Linke Seite: Der historische Meridianstein der Sternwarte wurde 2008 von dem Physiker Andreas Schrimpf wiederentdeckt und in seinem Auftrag 2010 originalgetreu saniert. Die Fläche des Steines beträgt etwa 1,50 x 1,50 m; die Breite der schwarzen Rechtecke 10 cm. Die Nordrichtung von der Westsäule der Sternwarte aus ist durch das fünfte Rechteck von rechts auf dem Stein gegeben.

Die Ausrüstung der Sternwarte erfolgte einem im 19. Jahrhundert weit verbreiteten Standard: Man nutzte damals die hohe Gleichförmigkeit der Drehung der Erde, um Positionen von Gestirnen durch den Zeitpunkt ihres Vorbeizugs (ihrer „Passage") an einer senkrechten Markierung im Teleskop zu bestimmen. Das Hauptinstrument war ein Passageinstrument, welches auf den Meridian (die Nord-Südrichtung) ausgerichtet werden musste. Es konnte in der Höhe verstellt werden, um die Passage von Sternen in verschiedenen Deklinationen zu vermessen. Gerling beschaffte 1840 ein Passageinstrument der Firma Ertel (München). Der zweite ebenso wichtige Teil der Ausrüstung war eine Präzisionsuhr, an der man die Durchgangszeit ablas. Diese Uhr erwarb Gerling bei der Firma Kessel (Altona), dem bedeutendsten Hersteller von Sternwartenuhren im 19. Jahrhundert. Eine solche Ausrüstung bot die Möglichkeit zu zweierlei Messungen: Zum einen konnten die Positionen von bisher unbekannten Objekten aus den gemessenen Durchgangszeiten ermittelt werden, zum anderen stellte das Vermessen der Durchgangszeiten bekannter Objekte eine präzise Bestimmung der lokalen Zeit dar.

Für diese Art der Messtechnik war eine drehbare Kuppel nicht erforderlich; es reichte eine feste, schwingungsarme Säule als Standort für das Teleskop, sowie eine freie Sicht vornehmlich nach Süden. Gerling errichtete mehrere Säulen in einer außen liegenden Galerie um ein kleines Beobachtungshaus auf der Spitze des Turms. Über einer der Säulen schützte ein bewegliches Dach das Passageinstrument.

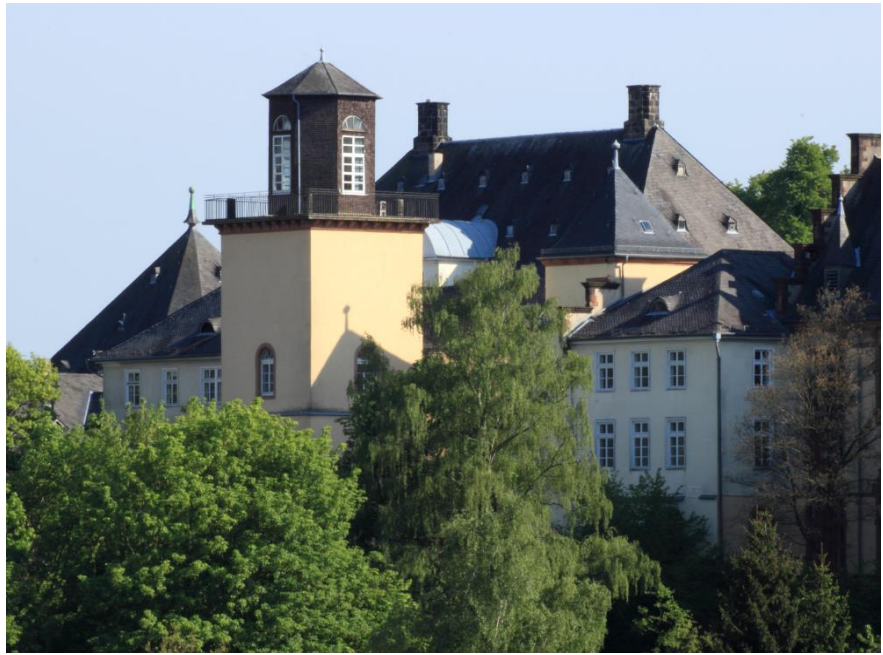

Die Gerling-Sternwarte befindet sich auf dem Turm des Gebäudes des ehemaligen Mathematisch-Physikalischen Instituts am Renthof 6.

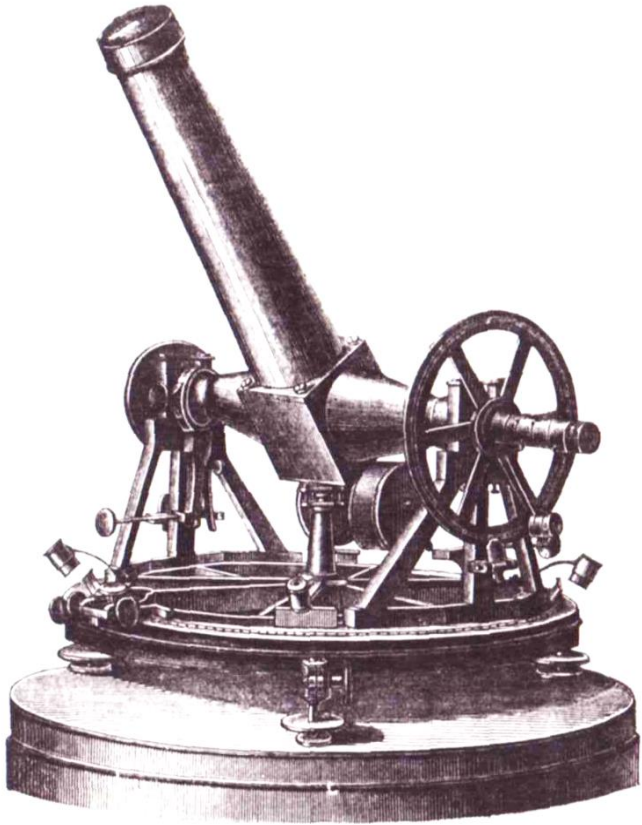

Das Passageinstrument kaufte Gerling 1840 bei der Firma Ertel (München). Es hat eine Objektivöffnung von 54 mm und eine Brennweite von 596 mm. Das Instrument besitzt eine „geknickte Optik": ein Umlenkspiegel im Kubus in der Mitte des Geräts erlaubt einen Einblick durch die Drehachse des Geräts. Die Beobachtung ist also unabhängig von der Höhe möglich, in die das Teleskop zeigt. Die Skizze ist dem Lehrbuch von Franz Melde entnommen.

Eine recht anspruchsvolle Vorarbeit für die Beobachtungen erfolgte durch die möglichst exakte Justierung des Fernrohrs auf den Meridian. Dazu mussten längere Messreihen von Sternpositionen um den Polarstern herum aufgenommen werden, anhand derer man dann die exakte Nordrichtung präzise ermitteln konnte. Um dies zu vereinfachen, setzte Gerling eine Landmarke, einen großen Stein mit einer aufgemalten Skala in knapp 4 km Entfernung in der Nordrichtung der Sternwarte. Man nennt eine Marke in dieser Position einen Meridianstein. Eine Einmessung der Skala ergab dann die exakte Lage der Nordrichtung des Passageinstruments. Zu Beginn einer Messnacht wurde die Skala beleuchtet und zunächst das Teleskop darauf ausgerichtet und justiert. Anschließend konnten mit dem justierten Teleskop Durchgangszeiten genau ermittelt werden.

Kataloge von Fundamentalsternen dienten dazu, die an verschiedenen Sternwarten gemessenen Durchgangszeiten relativ zueinander zu kalibrieren. Vermessen wurden lichtschwächere Sterne zur Verbesserung der Sternkarten, Planeten unseres Sonnensystems und deren Monde, sowie die Anfang des 19.

Jahrhunderts gerade entdeckten Kleinkörper des Sonnensystems im Asteroidengürtel. Gerlings Sternwarte ist die erste in Hessen, an der Positionsmessungen von Asteroiden (astrometrische Messungen) vorgenommen wurden. Diese historisch bedeutsamen Messungen waren 2011 der Anlass zur Taufe des Asteroiden 2008 CW 116 auf den Namen „Marburg". Noch heute sind die historischen Daten von Gerling und seinen Studenten in den Katalogen des Minor Planet Center enthalten und dienen der Verfolgung von Asteroiden-Bahnen.

Im Jahre 1862 setzte Gerling zwei weitere Justiersteine für das Fernrohr der Sternwarte: Den Oststein am Ortenberg und den Weststein in Wehrda. Damit konnte das Passageinstrument nun auch präzise nach Osten und Westen ausgerichtet werden, um Messungen im ersten Vertikal (also um 90 Grad gegen den Meridian verdreht) durchzuführen. Mit solchen Messungen kann die Höhe der Polachse über dem Horizont sehr genau vermessen und vor allem deren Änderung verfolgt werden.

Nach Gerlings Tod wurde die Sternwarte nicht mehr aktiv in der Astronomieforschung genutzt. Dennoch ist sie bis heute mehr oder weniger ununterbrochen in Benutzung von Studierenden und Mitgliedern des Fachbereichs Physik. Ein beredtes Zeugnis der Arbeitsweise einer Sternwarte im 19. Jahrhundert stammt von Gerlings Nachfolger Franz Melde (1832-1901), der die Marburger Erfahrungen in einem Lehrbuch festhielt.

Die historischen Geräte der Sternwarte sind leider alle verschollen, es gibt keinerlei Aufzeichnungen über deren Verbleib. Erhalten sind aber der Meridianstein und der Oststein, sowie die Säulen, die das Fernrohr trugen. Damit ist die Gerling-Sternwarte die einzige in Deutschland – vermutlich sogar in Europa – in der mit dem Meridianstein und dem Oststein Justiermarken noch an den originalen Plätzen erhalten und von der Sternwarte einsehbar sind. Dieses Observatorium stellt somit ein einzigartiges und wertvolles Zeugnis der Arbeitsweise der Astronomen im 19. Jahrhundert dar.

Bei Führungen, die der Fachbereich zu besonderen Gelegenheiten anbietet, wird die Marburger Astronomiegeschichte des 19. Jahrhunderts veranschaulicht.

*Andreas Schrimpf*

## Literatur

## Gerling Astronomical Observatory

Christian Ludwig Gerling's 1817 appointment as Professor for Mathematics, Physics and Astronomy resulted in the foundation of the Mathematical and Physical Institute of the Philipps University. In 1838, Gerling moved onto new premises in the main building of the former Dörnberger Hof in Renthof Street where the Philipps University's astronomical observatory was installed in the upper part of the old tower in 1841.

The most important device at that time was a transit instrument which served to measure the transit times of stars in the meridian. Precise alignment required the use of a meridian stone, an artificial point of reference exactly north of and at about four kilometers' distance from the observatory.

The scientists observed planets and their moons, the asteroids that were only discovered at the beginning of the 19th century, and some fainter stars in order to improve stellar charts. The Gerling Observatory is the first place in Hesse, where positions of asteroids were read!

**Gerling-Sternwarte**
Fachbereich Physik der Philipps-Universität Marburg
Renthof 6, 35037 Marburg
Ansprechpartner: Priv. Doz. Dr. Andreas Schrimpf
E-Mail: andreas.schrimpf@parallaxe-sternzeit.de
*http://www.parallaxe-sternzeit.de*

Besichtigung auf Anfrage. Sonderführungen werden auf der Internetseite bekannt gegeben.